\documentclass[aip,twocolumn,groupedaddress]{revtex4}
\usepackage{graphicx}
\usepackage{bm}
\usepackage{times}
\usepackage{amsmath}
\usepackage{float}

\begin{document} 



\title{X-ray reflectivity of chemically vapor deposited diamond single crystals in the Laue geometry \footnote{submitted for publication to Acta Cryst. A http://journals.iucr.org/a/}} 

\author{S. Stoupin}
\email{sstoupin@cornell.edu}

\author{J.P.C. Ruff}
\author{T. Krawczyk}
\author{K. D. Finkelstein}

\affiliation{Cornell High Energy Synchrotron Source, Cornell University, Ithaca, New York, USA}

\begin{abstract} 
Absolute X-ray reflectivity of chemically vapor deposited (CVD) diamond single crystals was measured in the Laue geometry in the double-crystal non-dispersive setting with an asymmetric Si beam conditioner crystal. The measurements were supplemented by rocking curve topography. The measured reflectivity curves are examined in the framework of Darwin-Hamilton approach using a set of two independent parameters, the characteristic thickness of mosaic blocks and their average angular misorientation. 
Due to strong extinction effects the width of reflectivity curves does not directly represent the average misorientation of the blocks. 
Two different sets of parameters were found for the 111 asymmetric reflection in the two different scattering configurations (beam compression and beam expansion). 
Analysis of the rocking curve topographs shows that this discrepancy can be attributed to inhomogeneity of the diamond crystal microstructure. 
\end{abstract}

\maketitle 



\section{Introduction}

Studies of X-ray reflectivity in imperfect crystals can be traced to the original work of Darwin \cite{Darwin22}, who introduced radiation intensity transfer equations considering X-ray absorption and incoherent superposition of X-rays scattered from misoriented blocks within a crystal.   
Further development was carried out by Hamilton \cite{Hamilton57} and Zachariasen \cite{Zach}. Solutions of the Darwin equations (later termed as Darwin-Hamilton equations) were analyzed and extinction effects were explored. The analytical solution for the general case of asymmetric diffraction in the transmission (Laue) geometry was reported by Dietrich and Als-Nielsen \cite{Dietrich65}. Generalized solutions in both Laue and Bragg geometries were reported and studied in more details by Sears \cite{Sears97-1,Sears97-2}. To date, diffraction theories of imperfect crystals have evolved substantially (see \cite{Authier} for a review). While the original interpretation of an imperfect crystal as an ensemble of uncorrelated misoriented blocks and the solutions of the Darwin-Hamilton equations can be considered as somewhat outdated, the model has found wide acceptance among neutron and X-ray communities as it permits a relatively simple interpretation and quantitative predictions, albeit with a limitation that direct evidence from an experiment is not readily available to either validate the theory or to show its limitations.

In this work, the theory is used to describe absolute reflectivity of chemically vapor deposited (CVD) single crystal diamond, 
a synthetic material with many emerging applications in modern technology. One of the promising applications of CVD diamond is their use 
as monochromators for neutrons, X-rays and gamma rays because the thermal and radiation hardness characteristics of diamond are superior to those of other single crystal materials. 
Distortions in the crystal lattice of a CVD diamond may enable selection of a greater fraction of incoming radiation from a polychromatic and/or divergent incident radiation compared to that of a perfect (or nearly perfect) single crystals. For a perfect crystal the fraction of radiation selected, and the resulting reflected flux are rather small due to the narrowness of the intrinsic angular-wavelength reflection region defined by the Darwin width. The case of transmission (Laue) geometry is studied here because it is favorable for high-heat-load monochromators operating at high X-ray photon energies ($E \gtrsim$~30~keV). 
The advantage of using the Laue geometry as opposed to the Bragg geometry stems from the ability to limit incident beam footprint on the entrance crystal surface for the most efficient low-index reflections (shallow Bragg angles). 
Reflectivity of CVD diamond has been studied earlier for applications in neutron monochromators \cite{Freund09,Freund11,Fischer13}. It was found that although reflectivity for neutrons can be close to the theoretical prediction the result strongly depends on variations 
in the diamond microstructure, which was deduced from X-ray reflectivity measurements. Our study is focused on absolute X-ray reflectivity, measured
in the non-dispersive double-crystal setting. Measurements were performed using a beam with 1$\times$1~mm$^2$ cross section and supplemented by rocking curve topography. It was found that for practical low-index reflections the approximation of ideally imperfect mosaic crystal is inapplicable. The effects of primary and secondary extinction were taken into account. Reasonable agreement with theory was obtained for the width of reflectivity curves and absolute reflectivity values. This was achieved by optimal choice of two independent parameters, the characteristic thickness of mosaic blocks and their average misorientation. It was also found that two different sets of parameters were required to describe reflectivities of an asymmetric reflection in the beam compressing and beam expanding geometries. This was attributed to inhomogeneous microstructure of the studied samples and the fact that X-ray-illuminated crystal volumes were inevitably different in these two geometries. The methodology developed in this work can be used for microstructural characterization and prediction of X-ray reflectivity of imperfect single crystals where the effects of primary and secondary extinction are substantial. It is not limited to diamond only. 

\section{Theory}\label{sec:theory}
\subsection{Reflecting power of a unit crystal thickness}\label{sec:theory1}
In the derivation of reflectivity a mosaic crystal in the form of plane parallel plate of thickness $T_0$ is considered \cite{Zach}. 
It is assumed that the thickness of individual crystal blocks fluctuates about a mean value $t_0$ which is small enough such that 
X-ray absorption within an individual block can be neglected. Various crystal blocks scatter independently of one another (i.e., uncorrelated model). 
The number of blocks that participate in scattering is large enough, that their angular misorientation can be represented by a continuous
function $W(\Delta$), where $\Delta$ is the angular deviation from the mean. For integration purposes it is assumed that a crystal 
slice with thickness $dT$ contains a large enough number of individual blocks. 

For a plane wave of photon energy $E_c$ the incident angle corresponding to the center of reflection region 
of a given $hkl$ reflection can be found from Bragg's law as described by the dynamical theory of X-ray diffraction:
\begin{equation}
E_H(1+w_H) = E_c \sin{\theta_c},
\label{eq:Bragg}
\end{equation}
where $E_H = hc/2d_{hkl}$ is termed Bragg energy, and $w_H$ is the refraction correction (e.g., \cite{Shvyd'ko_book}). 
For a plane wave of the same photon energy $E_c$ incident on the crystal at an arbitrary angle $\theta$ the reflecting power 
of a unit crystal thickness is
\begin{equation}
\sigma(\theta, E_c) = \frac{1}{t_0}\int^{\infty}_{-\infty} W(\Delta) R_0(\theta-\theta_c + \Delta, E_c) d \Delta ,
\label{eq:sigma1}
\end{equation}
where $R_0(\theta - \theta_c,E)$ is the zero-absorption solution for reflectivity of a single block (a perfect crystal plate with thickness $t_0$). 
Using substitution of variables $\tau = \theta - \theta_c + \Delta$, this integral can be interpreted as a convolution of the 
angular misorientation distribution $W(\Delta)$ and the reflectivity $R_0(\theta,E_c)$
\begin{equation}
\sigma(\theta, E_c) = \frac{1}{t_0}\int^{\infty}_{-\infty} W(\Delta \theta - \tau) R_0(\tau, E_c) d \tau,
\label{eq:sigma2}
\end{equation}
where $\Delta \theta = \theta - \theta_c$. 
In this work, the approximation of the reflectivity of the crystal block with a delta-function is not applied. 
Here, the effective width of the distribution function $W(\Delta)$ is allowed to be comparable to the angular width of $R_0(\tau,E_c)$ 
and integration in Eq.\ref{eq:sigma2} will be performed numerically. 
The main assumption about $R_0$ is that the interactions between the crystal blocks are not considered, 
their influence is reduced to a simple description given by the angular distribution function $W(\Delta)$, while variations in the local 
values of the lattice parameter are neglected. Far from the Bragg backscattering condition ($\theta < \pi/2$) the angular variations tend to dominate the relative variations in the lattice parameter (e.g., \cite{Macrander05,Tsoutsouva15}).
 
\subsection{Darwin-Hamilton equations and their solution in the Laue geometry} \label{sec:theory2}|
Now after the local reflecting power of a mosaic crystal has been defined we refer to the Darwin-Hamilton equations 
in the Laue geometry for the general case of an asymmetric reflection \cite{Dietrich65,Sears97-1}. 

\begin{eqnarray}
\frac{dI_0}{dT} = - \frac{\mu +\sigma}{\gamma_0} I_0 + \frac{\sigma}{\gamma_H} I_H, \\ \nonumber
\frac{dI_H}{dT} = - \frac{\mu +\sigma}{\gamma_H} I_H + \frac{\sigma}{\gamma_0} I_0  
\label{eq:DH}
\end{eqnarray}
Here, $I_0$ is the intensity of the transmitted wave, $I_H$ is the intensity of the reflected wave, $\mu$ is the X-ray absorption coefficient, $\gamma_0$ and $\gamma_H$ are the direction cosines for the transmitted and reflected directions respectively.
Using boundary conditions for the Laue geometry ($I_H(0)$~=~0 and $I_0(0) = I_0$) the reflectivity solution of these coupled differential equations is
\begin{equation}
R(\theta,E_c) = \frac{I_H(T_0)}{I_0} = \frac{\sigma}{\gamma_0 \varepsilon} \exp{\bigg(- \frac{\mu +\sigma}{\Gamma}T_0 \bigg)} \sinh{(\varepsilon T_0)}
\label{eq:DHsol}
\end{equation}
where 
\begin{equation}
\varepsilon = \sqrt{\frac{(\mu + \sigma)^2}{G^2} +\frac{\sigma^2}{\gamma_0 \gamma_H}},
\end{equation}
and 
\begin{equation}
\frac{1}{\Gamma} = \frac{1}{2}\bigg(\frac{1}{\gamma_0} + \frac{1}{\gamma_H} \bigg ); 
\frac{1}{G} = \frac{1}{2}\bigg(\frac{1}{\gamma_0} - \frac{1}{\gamma_H} \bigg ).
\end{equation}

The angular variations of the direction cosines $\gamma_0$ and $\gamma_H$ in vicinity of $\theta = \theta_c$
can be neglected due to smallness of the angular range of reflection of any crystal block 
and the width of the angular distribution function $W(\Delta)$. 
Thus, the angular and photon energy dependencies in Eq.~\ref{eq:DHsol} are due to $\sigma(\theta, E_c)$ and $\mu(E_c)$. 
It is also noted that for the low order reflections in diamond the thickness of crystal blocks $t_0$ can be comparable to the 
Pendell\"{o}sung distance and that for hard X-rays the absorption coefficient is small compared to the scattering power ($\mu \ll \sigma$). 
Therefore, the effects of primary and secondary extinction become substantial. Thus, the approximation of "ideal mosaic crystal" 
where the opposite is true (see e.g., \cite{Zach}) is not applicable.  
\subsection{Optimal plate thickness. Maximum attainable reflectivity.} \label{sec:theory3}|
For a given Laue reflection there exists an optimal thickness of the mosaic crystal plate
for which reflectivity is maximized \cite{Sears97-1}. Differentiation of Eq.\ref{eq:DHsol} yields
\begin{equation}
T_{opt} = \frac{1}{\varepsilon} \tanh^{-1}{\bigg(\frac{\varepsilon \Gamma}{\mu + \sigma}\bigg)}
\label{eq:Topt}
\end{equation}

Here, the values of $\varepsilon$ and $\sigma$ are assumed to be taken at the center of reflection region $\theta = \theta_c$.

The smallness of X-ray absorption in diamond permits interesting observations on the maximum possible values of reflectivity. 
If $\mu = 0$ the reflectivity solution can be simplified as follows \cite{Sears97-1}.

\begin{equation}
R = \frac{1 - \exp{[-\Sigma (1 + b_H)]}}{1+b_H},
\label{eq:DHsol0}
\end{equation}

where $\Sigma = \sigma T_0/\gamma_0$, and $b_H = \gamma_0/\gamma_H$ is the 
asymmetry factor of the reflection. From Eq.~\ref{eq:DHsol0} it becomes clear that reflectivity of a symmetric Laue reflection ($b_H = 1$) can
not be greater than 1/2. It can become greater than 1/2 in the beam expanding geometry $b_H < 1$. This situation however, does not lead to
increase in the radiation flux density [photons/cm$^2$] because the change in the beam size upon reflection is proportional to $b_H$. 
For practical considerations the ratio of radiation flux density upon an asymmetric reflection to that of a symmetric reflection is considered,
which is termed Fankuchen gain (see \cite{Sears97-1} and references therein). From Eq.~\ref{eq:DHsol0} 

\begin{equation}
G = \frac{2b_H}{1+b_H}  \frac{1 - \exp{[-\Sigma (1 + b_H)]}}{1 - \exp{[-2 \Sigma]}},
\label{eq:Fchen}
\end{equation}
Thus, gain in the flux density can be obtained using an asymmetric reflection in the beam compressing geometry ($b_H > 1$).    
\subsection{Blocks formed by dislocation walls. Numerical example of the symmetric 220 Laue reflection.}\label{sec:theory4}
A simple microstructural model of a mosaic crystal was proposed by Burgers \cite{Burgers40}. 
In this model the boundary surfaces between the differently oriented blocks (or domains) 
are formed by sets of parallel dislocation lines 
located at equal distance $h$ apart. The relative misorientation angle of neighboring domain walls is $\alpha = b/h$, 
where $b$ is the modulus of the dislocation's Burgers vector. If $h_0$ is the lateral size of the block containing $h_0/h$ dislocations, the 
dislocation density is $\rho = \alpha/(b h_0)$ \cite{Gay53}.   

In general, the thickness of the block $t_0$ and the mosaic spread $\alpha$ are two physically independent parameters, which means that 
the lateral size of the block $h_0$ is not related to $t_0$.   
It is instructive, however, to consider a regular grid of dislocation such that 
\begin{eqnarray}
\label{eq:iso}
h = h_0 = t_0, \\ \nonumber
\alpha = b/h_0 = b/t_0.
\end{eqnarray}
In this case the misorientation angle is inversely proportional to the block thickness.

It is assumed that the dislocation lines are of edge type along the [001] direction with magnitude of Burgers vector $b = a \sqrt{2}/2$, 
where $a$ is the lattice parameter. The angular misorientation function is approximated with a Gaussian distribution
\begin{equation}
W(\Delta) = \frac{1}{\sqrt{2\pi}\Delta_0}\exp{\bigg(- \frac{\Delta^2}{2\Delta_0^2} \bigg)},
\label{eq:W}
\end{equation}
where a rather small width is chosen $\Delta_0 \simeq 10 \mu$rad to limit total angular acceptance range of the reflection, 
thus limiting the possible angular spread of the reflected radiation.  
Equation \ref{eq:iso} yields $t_0 \simeq 11 \mu$m. For such a simplified system we calculate reflectivity 
curves of a symmetric 220 Laue reflection ($b_H = 1$) at different photon energies for an optimal thickness of the diamond plate 
(Eq.~\ref{eq:Topt}) at each photon energy. 
In calculations of the reflecting power of a unit crystal thickness (Eq.~\ref{eq:sigma2})
we use  zero-absorption solution for reflectivity $R_0(\theta,E_c)$ of a perfect diamond crystal 
given by the dynamical theory of X-ray diffraction (e.g., \cite{Shvyd'ko_book}). The polarization factor of the incident X-ray wave 
is assumed to be unity (e.g., $\sigma$-polarization). 

\begin{figure*}
\caption{Calculated reflectivities (as functions of the angular deviation from the center of the reflection region) 
of the symmetric 220 Laue reflection in diamond at different photon energies (20, 40 and 80 keV) for a diamond plate of optimal thickness (Eq.~\ref{eq:Topt}). 
The reflectivities for the mosaic plate (described by Eq.~\ref{eq:DHsol}) are shown with dashed blue lines. 
Reflectivities of a perfect crystal plate of the same thickness are plotted with solid red lines.
The angular misorientation function $W(\Delta)$ with a standard deviation $\Delta_0$~=~10~$\mu$rad normalized to its maximum value is shown
in each subplot as a reference for curve width (black solid line).} 
\vspace*{\floatsep}
\includegraphics[scale=0.89]{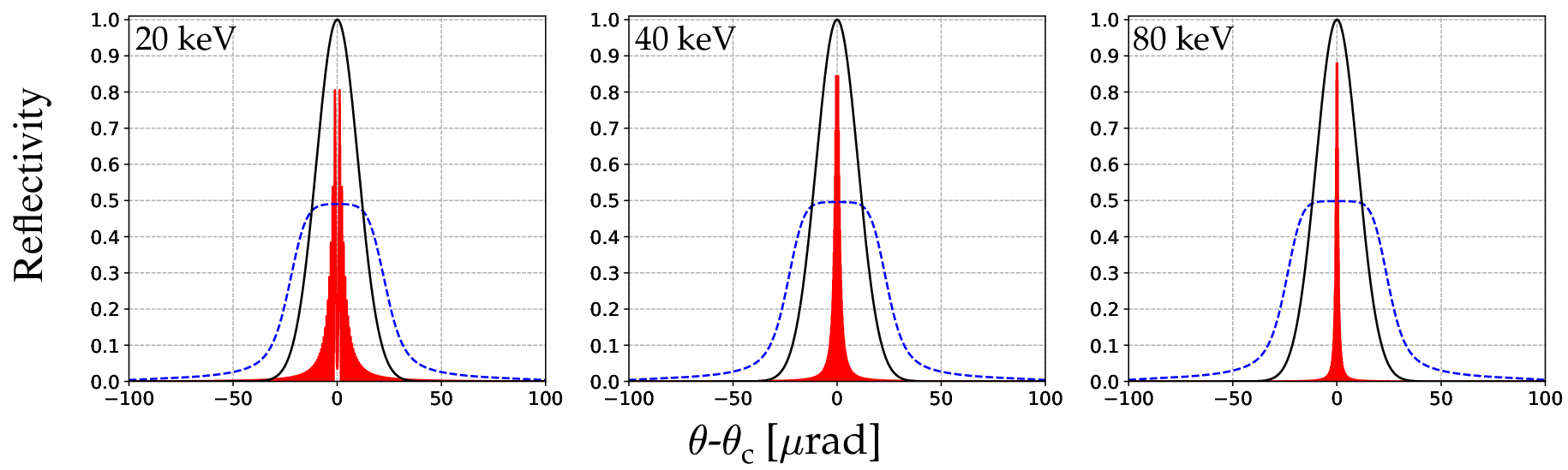} 
\label{fig:refl_220}
\end{figure*}

Figure~\ref{fig:refl_220} shows the calculated reflectivities as functions of the angular deviation of the monochromatic X-ray 
from the center of the reflection region. The three subplots show calculations at photon energies of 20, 40 and 80 keV. 
The calculated reflectivities for the mosaic plate (Eq.~\ref{eq:DHsol}) are shown with dashed blue lines. 
For comparison, reflectivities of a perfect crystal plate of the same thickness ($T_0 = T_{opt}$) are plotted with solid red lines. 
These exhibit fast oscillations due to interference between the X-ray wavefields propagating in the crystal. The angular misorientation
function $W(\Delta)$ normalized to its maximum value is shown in each subplot (black solid line) to serve as a reference for curve width. 
The reflectivities of the mosaic plate exhibit a flat top portion. The effect can be understood
based on Eq.~\ref{eq:DHsol0}, where the scattering factor $\Sigma(1+b_H)$ under the exponent attains values greater than unity and 
the angular width of the reflectivity function becomes greater than the width of the misorientation function $W(\Delta)$.
The physical interpretation of this phenomenon is the effect of secondary extinction: the X-ray wavefields propagating within the crystal
exhibit many scattering interactions with individual blocks. If X-ray absorption is negligibly small
and the number of these events is sufficiently large the intensity of the incident wave (i.e., radiation flux density [photons/$cm^2$]) 
is equally split between the transmitted and the reflected branches at the exit surface.
It is of interest to compare the integrated reflectivity of the mosaic crystal to that of a perfect crystal at different photon energies.
These ratios along with other parameters of the calculations are shown in Table~\ref{tab:param220}.
As expected, the ratio of the integrated reflectivities increases with photon energy because the intrinsic angular width 
of the reflection in the perfect crystal decreases dramatically while the width of the reflectivity curve of the mosaic crystal remains 
practically unchanged. 

\begin{table} 
\label{tab:param220}
\caption  {Parameters of calculations of the 220 Laue reflectivity curves at different photon energies.\\
$T_{opt}$ - optimal thickness of the crystal plate (Eq.~\ref{eq:Topt}), \\
$2\times \Sigma$ - total scattering factor (Eq.~\ref{eq:DHsol0}), \\
$\Delta \theta^{p}$ - reflectivity curve width for the perfect crystal plate (FWHM), \\
$\Delta \theta^{m}$ - reflectivity curve width of the mosaic crystal plate (FWHM),   \\
$R^{m}_{int}/R^{p}_{int}$ - ratio of integrated reflectivities of the mosaic and perfect crystal plates.\\
} 
\begin{tabular}{l c c c}  
\hline\hline               %
Photon energy [keV]                      &  20            &  40            & 80     \\
$T_{opt}$ [mm]                           &  0.21          &  0.93          & 4.2    \\
$2\times \Sigma$                         &  5.9           &  6.8           & 7.8    \\
$\Delta \theta^{p}$~[$\mu$rad]           &  4.9           &  2.4           & 1.2    \\
$\Delta \theta^{m}$~[$\mu$rad]           &  48            &  49            & 51     \\ 
$R^{m}_{int}/R^{p}_{int}$                &  7.6           &  16.2          & 31.5   \\
\hline\hline
\end{tabular} 
\end{table} 

\section{Crystal plates and their preliminary characterization.}\label{sec:samples}
Four rectangular CVD diamond crystal plates with (001) surface and (110) side orientation were obtained from different crystal manufacturers.  
Also, two rectangular plates synthesized by the high-pressure high-temperature method (HPHT) were used in reference reflectivity measurements. 
Characteristics of the plates are summarized in Table~\ref{tab:sample}.

\begin{table*} 
\label{tab:sample}
\caption  {Characteristics of studied diamond crystal plates.} 
\begin{tabular}{l c c c c c c}               
\hline\hline
Diamond plate                            &  CVD-1         &  CVD-2          & CVD-3             & CVD-4          & HPHT-1         & HPHT-2 \\
Synthesis method                         &  CVD           &  CVD            & CVD               & CVD            & HPHT           & HPHT   \\
Manufacturer's grade                     & Optical        &  Optical        & Optical           & Electronic     & IIa            & IIa    \\
Dimensions~[mm]~($L_0 \times W_0$) & 7.2$\times$5.8 & 7.0$\times$6.0  & 7.0$\times$3.5    & 8.2$\times$6.1 & 4.0$\times$4.0 & 4.0$\times$4.0 \\
Thickness~[mm]  ($T_0$)            & 0.64           & 1.02            & 0.68              & 0.54           & 0.55           & 0.55         \\                
\hline\hline
\end{tabular} 
\end{table*} 

\begin{figure}
\caption{White-beam X-ray Laue topographs of selected CVD crystal plates:
(a)  CVD-3 (optical grade); the image is obtained from $3\bar{1}1$ reflection. 
Digital enhancement of the nearly saturated image reveals contrast due to a dense dislocation network. 
(b)  CVD-4 (electronic grade); the image is obtained from $131$ reflection. 
} 
\includegraphics[scale=0.4]{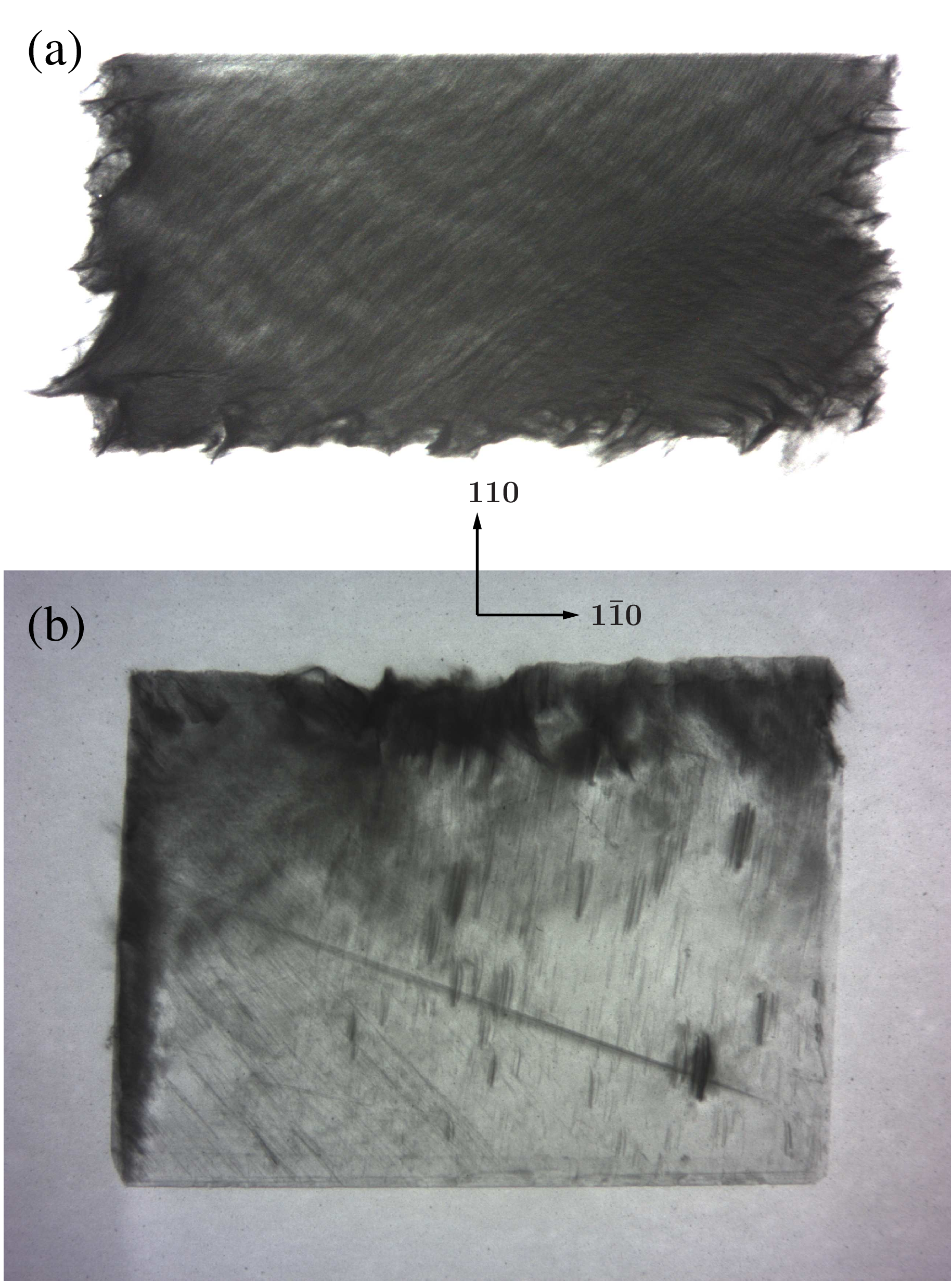} 
\label{fig:wbxt}
\end{figure}

White-beam X-ray topography in the transmission geometry was performed for several plates at 1BM beamline of the 
Advanced Photon Source \cite{Stoupin_AIPP16_1,Macrander15}. 
Major differences were observed for CVD plates of optical grade compared to plates of electronic grade. 
Figure \ref{fig:wbxt}(a) shows white beam topograph obtained from $3\bar{1}1$ reflection of CVD-3 (optical grade) diamond plate and Fig.~\ref{fig:wbxt}(b) 
shows white beam topograph obtained from $131$ reflection of CVD-4 (electronic grade) diamond plate. 

While the specific details of crystal growth conditions remain unknown, during the synthesis of electronic grade plates chemical precursors 
with higher purity and substrates with higher quality (IIa type HPHT diamond plates) are used to minimize density of electrically active defects. 
The resulting overall crystal quality is typically better, which is confirmed by the white-beam topographs. While the X-ray exposure time was 
the same, the intensity of reflected X-rays was much greater for plate CVD-3. The reflection nearly saturated the X-ray film. Using image enhancement 
techniques applied to the topograph the contrast variation due to bundles of dislocations can still be seen (Fig. \ref{fig:wbxt}(a)). 
At the same time, the electronic grade plate CVD-4 reveals very inhomogeneous distribution of crystal lattice defects. 
Defect-free small-volume regions are observed along with strongly distorted regions near the top edge. The topograph of plate CVD-3 suggests 
that crystals with dense distribution of defects can be more efficient as X-ray reflectors. 

Another preliminary observation is the fact that the topograph of plate CVD-3 reveals substantial distortion clearly seen at the bottom, 
left and right edges of the image which suggests strong distortion of the crystal lattice, 
possibly due to non-uniform conditions during diamond synthesis (e.g., temperature distribution). 
The absence of such features at the upper edge was found to be due to the fact that the plate originated from a larger crystal 
which was laser-cut in two equal parts. The cut apparently corresponds to the upper edge in the topograph. 

In a broader sense, use of imperfect crystals as X-ray reflectors inevitably leads to distortion of radiation wavefront due to local deviations from 
the X-ray diffraction condition. However, greater integrated reflectivity of an efficient mosaic reflector can outweigh this deteriorating effect 
of wavefront distortion on the reflected flux density. 
To explore this possibility measurements more quantitative in nature than the conventional white-beam X-ray topography are required.
Below we describe a double-crystal setup that permits measurements of absolute reflectivity in the non-dispersive configuration 
and simultaneous measurement of rocking curve topographs of the probed crystal region. 

\section{Experiment}\label{sec:exp}
\begin{figure*}
\caption{Experimental setup. Synchrotron X-ray beam produced by bending magnet source (not shown) passing through the double-crystal 
Si (220) monochromator (DCM, symmetric reflections) and through an adjustable aperture (SLITS) is incident on the Si beam conditioner 
crystal (asymmetric (220) reflection). 
The intensity of the expanded beam reflected from the beam conditioner is measured using ionization chamber IC1. The intensity of the beam 
reflected from the diamond crystal plate is measured using ionization chamber IC2 while its profile is imaged using an area detector (AD). 
For each studied crystal plates measurements were performed at two different orientations of the diamond crystal plate:
(a)  Laue reflection from the ($\bar{1}\bar{1}\bar{1}$) planes in the beam expanding geometry,  
(b)  Laue reflection from the ($\bar{1}\bar{1}1$) planes in the beam compressing geometry
The portion of the reflected beam which can be traced to diffraction of the transmitted wave from the bulk of the crystal 
is marked with light red color. This portion contained inside the crystal corresponds to the Borrmann triangle, 
which illustrates the effect of X-ray propagation and reflection in the crystal with finite thickness.}
\vspace*{\floatsep}
\includegraphics[scale=0.89]{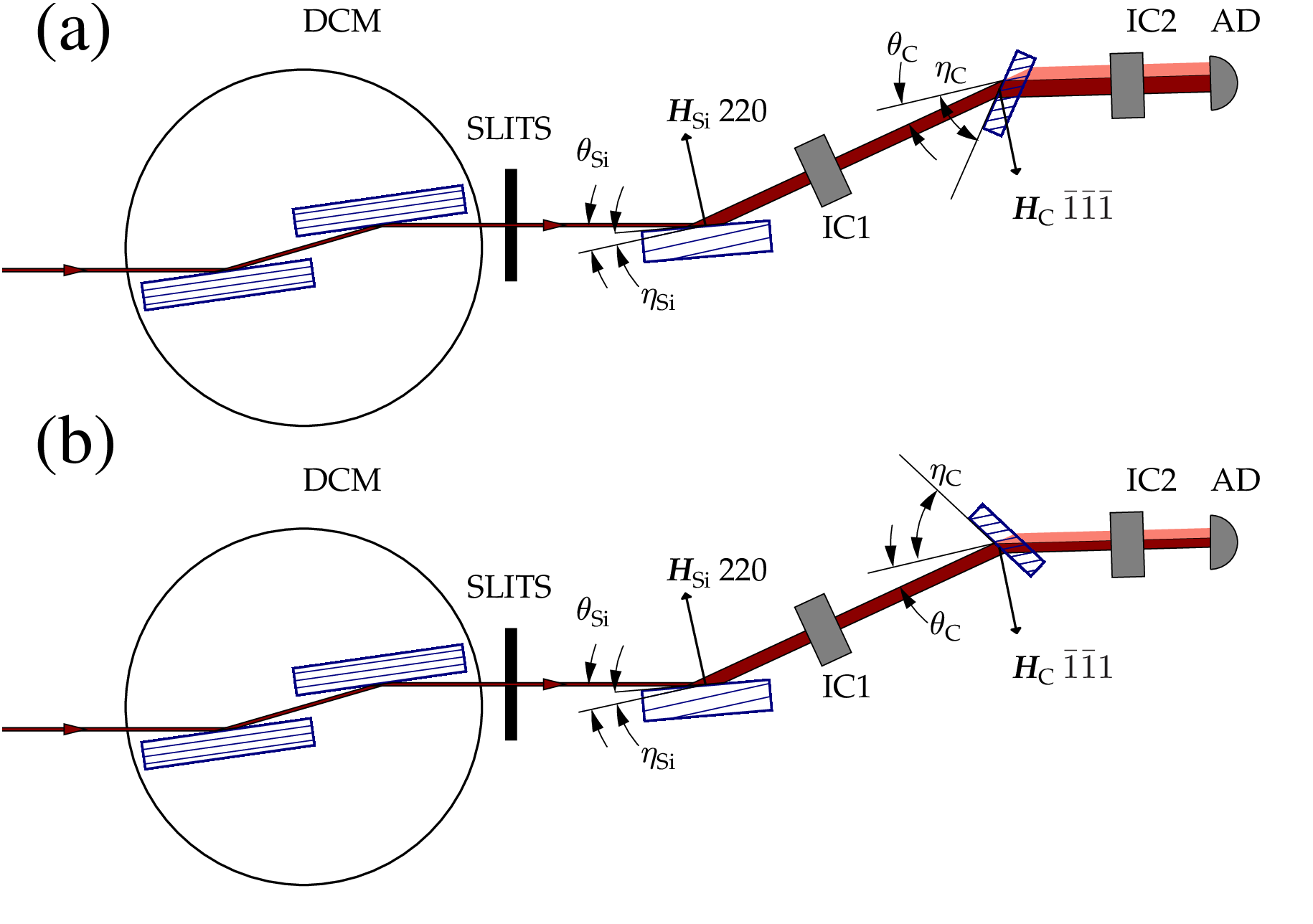} 
\label{fig:setup}
\end{figure*}

Preliminary measurements were performed at 1-BM beamline of the Advanced Photon Source using rocking curve X-ray topography capabilities \cite{Stoupin_AIPP16_2}. 
Only a subset of samples was studied using this setup at a photon energy of 8 keV. 
Simultaneous measurements of the beam intensities with calibrated detectors was not performed. 
Thus, the absolute reflectivity values could only be deduced indirectly (e.g., in relation to a known standard).
The results of these preliminary measurements served as guidance for further experimentation.
The final experiment was conducted at C1 beamline of Cornell High Energy Synchrotron Source using a similar setup with an addition of ionization chambers 
to simultaneously monitor intensities of the beam incident on the diamond crystal plate and the beam reflected from the plate in the Laue geometry
as shown in Fig.~\ref{fig:setup}. Synchrotron X-ray beam generated by bending magnet source (not shown) was passing through the double-crystal 
Si (220) monochromator (DCM, symmetric reflections) tuned to a photon energy of 15 keV. 
The monochromatized beam apertured using adjustable X-ray slits was incident on an asymmetric Si crystal (asymmetric (220) reflection). 
The intensity of the beam reflected from the asymmetric Si crystal was monitored using an ionization chamber IC1 set in the path of the beam. 
The reflected beam was incident on the diamond crystal plate set in the non-dispersive configuration for the 111 Laue reflection.
The reflection parameters are given in the supporting information.  
 
The intensity of the beam reflected from the diamond plate was monitored using an ionization chamber IC2. 
The beam transmitted through this ionization chamber was imaged using an area detector AD. Detectors IC2 and AD were conveniently placed 
on 2$\theta$ arm of a six-circle Huber goniometer, while the diamond plate was mounted at its center. This arrangement enabled calibration 
of the two ionization chambers to each other at the 2$\theta$ setting corresponding to zero scattering angle. 
A single calibration measurement took into account slightly different sensitivities of the ionization chambers, 
as well as X-ray absorption in air and the sample holder material (1-mm-thick supporting surface made of polyimide). 

Measurements were performed while scanning the angle of the diamond crystal plate in the scattering plane ($\theta$) 
over the 111 reflection curve while taking snapshots of the beam profile at each angular setting of the crystal. 
Rocking curve topographs were computed from the resulting sequences of images. The size of the beam reflected 
from the beam conditioner Si crystal was limited to a cross section of 1$\times$1~$mm^2$ using slits placed upstream of the crystal. 
In addition, a sequence of images was taken with wide open slits (with diamond plate fully illuminated by the beam) to test inhomogeneity 
of the whole plate with rocking curve topography. For each studied plate measurements were performed at two different orientations of the 
crystal plate, which correspond to beam expanding geometry (Fig.~\ref{fig:setup}(a)) and beam compressing geometry (Fig.~\ref{fig:setup}(b)). 
Indexing of the reciprocal vectors of the diamond crystal is consistent with the orientation of the plates as shown in Fig.~\ref{fig:wbxt}.
Prior to experimental studies of CVD plates reference measurements on HPHT plates were performed. The results were found to be in a 
reasonable agreement with predictions of dynamical theory of X-ray diffraction for perfect crystals (see the supporting information for details). 

\section{Results and Discussion}\label{sec:results}
\begin{figure*}
\label{fig:refl_15keV}
\caption{Reflectivities of CVD diamond plates measured in the Laue geometry (Fig.~\ref{fig:setup}). 
The experimental points measured in the beam expanding geometry (Fig.~\ref{fig:setup}a) are shown with squares(green) 
while those measured in the beam compressing geometry (Fig.~\ref{fig:setup}b) are shown with circles (blue). 
The green and blue solid lines represent theoretical reflectivity curves of beam expanding and beam compressing geometry respectively. 
Matching the theoretical curves to the experimental data was achieved by optimal adjustment of the thickness of mosaic block 
$t_0$ and the standard deviation of angular misorientation $\Delta_0$ (Table~\ref{tab:rc}).
}
\includegraphics[scale=0.95]{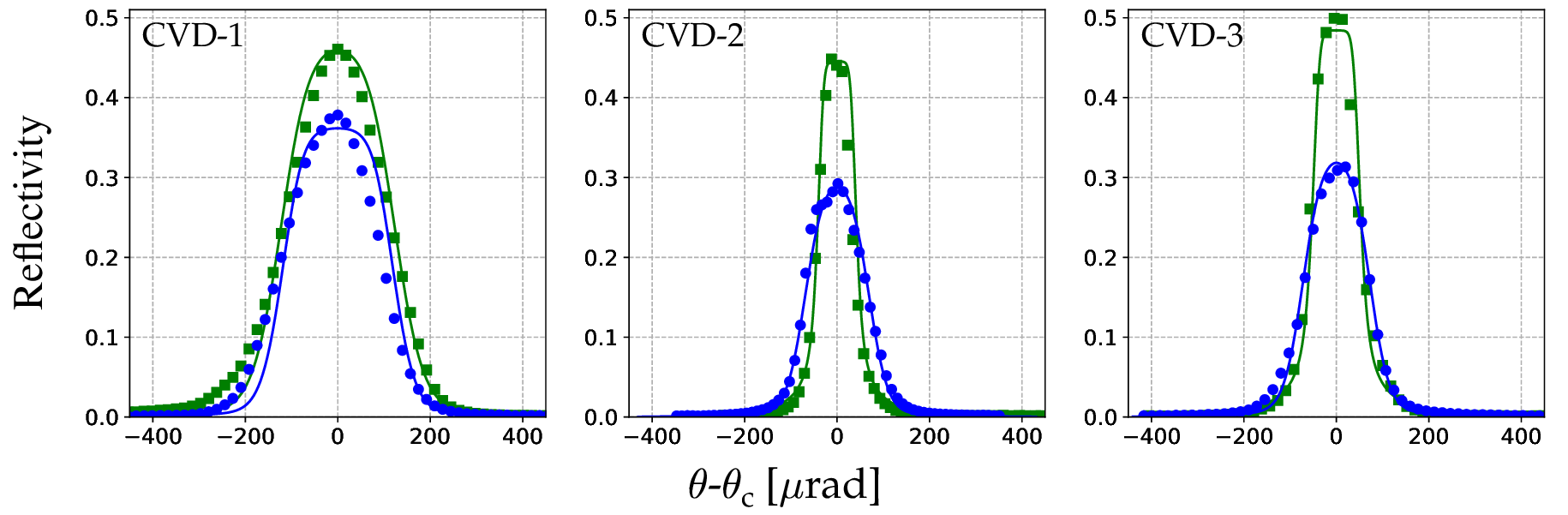} 
\end{figure*}

Table~\ref{tab:rc} summarizes the results of reflectivity measurements with the incident beam limited to a size of 1$\times$1~mm$^2$. 
The beam footprint was approximately centered on the entrance surface of the studied diamond plates. 
Figure~\ref{fig:refl_15keV} shows reflectivity curves for each plate. The experimental points measured in the beam expanding geometry 
are shown with squares(green) while those of the beam compressing geometry are shown with circles (blue). 
The experimental uncertainties estimated using fluctuations of detector readings (IC1 and IC2) were found to be 
slightly less than the size of the markers. The green and blue solid lines represent theoretical reflectivity curves matching the experimental data 
using optimally adjusted thickness of mosaic block $t_0$ and the standard deviation of angular misorientation $\Delta_0$. 
These parameters are listed in Table~\ref{tab:rc} along with the total width of the reflectivity curve $\Delta \theta$ (full width at half maximum or FWHM)
for each plate and the two reflection geometries. The values corresponding to beam compressing geometry are given in parentheses. 
For each case the total width of the reflectivity curve is noticeably greater compared to the effective angular misorientation $\Delta_0$
when the latter is scaled to represent FWHM. This observation suggests strong secondary extinction in diamond as was discussed in 
detail in section~\ref{sec:theory4}. 
      
\begin{table} 
\label{tab:rc}
\caption  {Characteristics of the probed regions for the studied CVD diamond plates.
The regions were illuminated by 1$\times$1~mm$^2$ incident beam and approximately centered on the diamond plates.  
The measured width of the curves (FWHM) is denoted by $\Delta \theta$. 
The thickness of the mosaic block $t_0$ and the standard deviation of the angular misorientation $\Delta_0$ 
are derived by matching theoretical and experimental reflectivity curves for the 111 Laue reflection.
The integrated reflectivity is compared to that of the perfect crystal of the same thickness ($R^{m}_{int}/R^{p}_{int}$). 
For each parameter two values are given corresponding to the beam expanding ($b_H < 1$) and the beam compressing ($b_H > 1$) geometries. 
The values in parentheses correspond to the beam compressing geometry. \\
} 
\begin{tabular}{l c c c c}           
\hline\hline
Diamond plate                            &  CVD-1         &  CVD-2          & CVD-3             & CVD-4      \\
$\Delta \theta$ (FWHM) [$\mu$rad]        &  242 (226)      &  78 (142)        & 107 (148)          & 28 (26)      \\
$\Delta_0$ (r.m.s.) [$\mu$rad]           &   70 (60)       &  16 (43)         & 20 (44)            & 6 (5)        \\
$t_0$     [$\mu$m]                       &   25 (18)       &  55 (65)         & 25 (30)            & 282 (149)   \\
$R^{m}_{int}/R^{p}_{int}$                &   12.6 (12.7)   &  5.9 (7.4)       & 7.3 (8.1)          & 1.4 (1.6)    \\
\hline\hline
\end{tabular} 
\end{table} 

The obtained values of $t_0$ and $\Delta_0$ are on the same order of magnitude as those considered in Section~\ref{sec:theory4} derived as estimates 
based on simple assumptions about the microstructure of mosaic crystal. Aside from the main goal of predicting X-ray reflectivity based 
on known crystal microstructure one can attempt to solve the inverse problem, i.e., to gain insight on the microstructure based on the results of 
reflectivity measurements. 
First, the obtained structural parameters need to be validated based on the limitations of the approach. 
For plate CVD-4 the thickness of the mosaic block seems to be quite large, comparable with the thickness of the crystal $T_0$. 
This contradicts the assumptions on $W(\Delta)$ being a continuous function. 
In such scenario, describing radiation transfer using the differential equations is inappropriate because the number of blocks 
contributing to X-ray scattering is limited. The white beam topograph of CVD-4 suggests that this is indeed the case, since certain regions 
of the crystal appear to be nearly perfect. 
Thus, the parameters derived for CVD-4 can be interpreted only as semi-quantitative indicators. The reflectivity curves for CVD-4 are 
presented in the supporting information. 

The model of the mosaic crystal used in this work assumes that characteristics of the blocks are homogeneous across the studied crystal region. 
In this regard, rocking curve topography (originated by \cite{Lubbert00} and also commonly referred to as rocking curve imaging) offers additional insights. 
We note that the rocking curve topographs of the diamond plates in the Laue geometry represent projections of the crystal volume across its entire thickness 
in the observation plane (e.g., \cite{TranThi17}), just like in any other X-ray diffraction topography technique applied in the Laue geometry. 
Unlike conventional X-ray topography (using either white-beam or monochromatic X-rays) where quantitative analysis is focused on studies of defect-induced diffraction contrast (e.g., \cite{Lang83,Moore09}) rocking curve topography offers quantitative mapping of macroscopic characteristics (e.g., reflectivity, lattice tilt, curve's width). 

\begin{figure*}
\label{fig:rct_c111}
\caption{Rocking curve topographs of plate CVD-1 in the beam expanding geometry:
(a)  for the entire crystal plate; the region probed with the 1$\times$1 mm$^2$ incident beam for absolute reflectivity is marked with a dashed rectangle,
and, (b) for the probed region. 
In each case the topographs represent maps of the reflected intensity at the curve's peak value normalized by the maximum value observed ($I^{peak}_R$), 
the curve's peak position ($\delta \theta_m$) and the curve's width as a standard deviation of a Gaussian approximation ($\Delta \theta_{\sigma}$).
The colorbar on the $\delta \theta_m$ and $\Delta \theta_{\sigma}$ topographs are in units of $\mu$rad. 
} 
\includegraphics[scale=0.8]{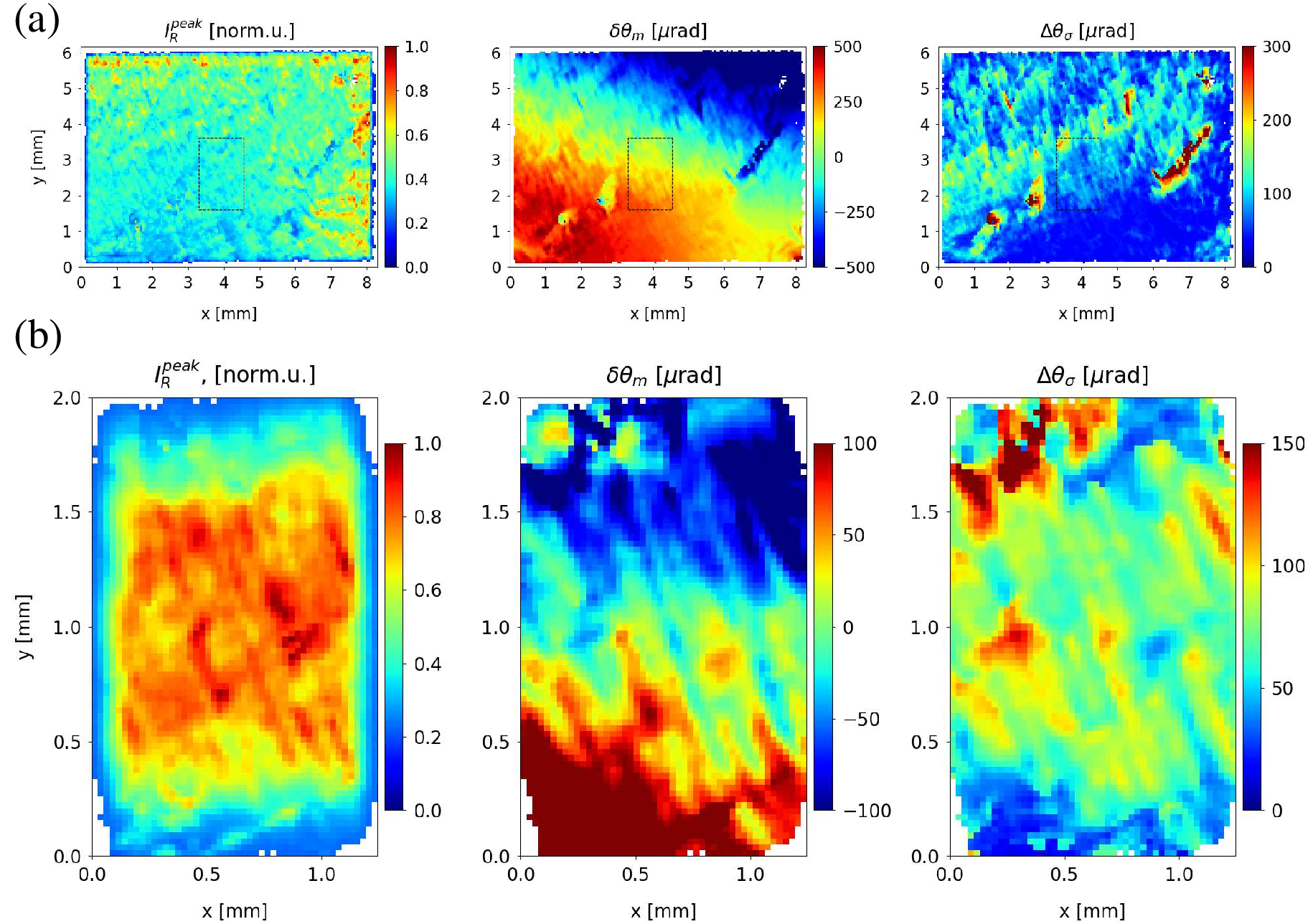} 
\end{figure*}

From Table \ref{tab:rc} the total width of the rocking curve for plate CVD-1 seems to be rather large ($\Delta \theta$~=~242$\mu$rad) if compared with 
those of the other two plates of the same optical grade (CVD-2 and CVD-3). Also, for CVD-1 a more significant mismatch in the shape of the 
reflectivity curves is observed (particularly on the tails).  
The topographs for the entire crystal plate CVD-1 are shown in Fig.~\ref{fig:rct_c111}(a) representing maps of 
reflected intensity (peak value) normalized by the maximum value observed ($I^{peak}_R$), the curve's peak position ($\delta \theta_m$) 
and curve's width as a standard deviation of a Gaussian approximation ($\Delta \theta_{\sigma}$). For brevity, only topographs corresponding 
to the beam expanding geometry are shown. The region probed with the 1$\times$1 mm$^2$ incident beam for absolute reflectivity is marked with a dashed rectangle.
The topographs corresponding to the probed region are shown in Fig.~\ref{fig:rct_c111}(b).
While the locally reflected intensity appears to be homogeneously distributed within the probed region, the peak position exhibits systematic variation across the entire crystal and within the region. This effect indicates the presence of a systematic overall curvature of the diffracting planes and contributes to the 
total width of the reflectivity curve in the probed region. Indeed the width of the rocking curve averaged across the region 
($\Delta \theta_m$~=~176~$\mu$rad) was found to be less than the total width ($\Delta \theta$~=~242$\mu$rad). 


Another observation from Table~\ref{tab:rc} is the mismatch of the values corresponding to the beam compressing and beam expanding geometry for each plate. 
There are two possible explanations. First, different volumes of the crystal are probed in the two geometries as illustrated in Fig.~\ref{fig:setup}
with the larger volume due to the shallower incidence angle $\eta_C - \theta_C$ in the beam expanding geometry. Therefore, different sets of values are expected 
if distribution of defects is inhomogeneous. Second, two different set of planes are probed ($\bar{1}\bar{1}\bar{1}$) and ($\bar{1}\bar{1}1$). 
These are crystallographically equivalent in a perfect diamond crystal. However, in a mosaic crystal it is expected that the crystal blocks are not necessarily 
constrained to be isotropic with respect to the crystal axes \cite{Sabine-ITC-C}. 

\begin{figure*}
\label{fig:rct_cvda}
\caption{Rocking curve topographs of plate CVD-3 (entire crystal) in the beam expanding geometry (a) and in the beam
compressing geometry (b) showing maps of the reflected intensity at the curve's peak value normalized by the maximum value observed ($I^{peak}_R$), 
the curve's peak position ($\delta \theta_m$) and the curve's width as a standard deviation of a Gaussian approximation ($\Delta \theta_{\sigma}$).
The region probed with the 1$\times$1 mm$^2$ incident beam for absolute reflectivity is marked with a dashed rectangle.
The colorbar on the $\delta \theta_m$ and $\Delta \theta_{\sigma}$ topographs are in units of $\mu$rad.
} 
\includegraphics[scale=0.89]{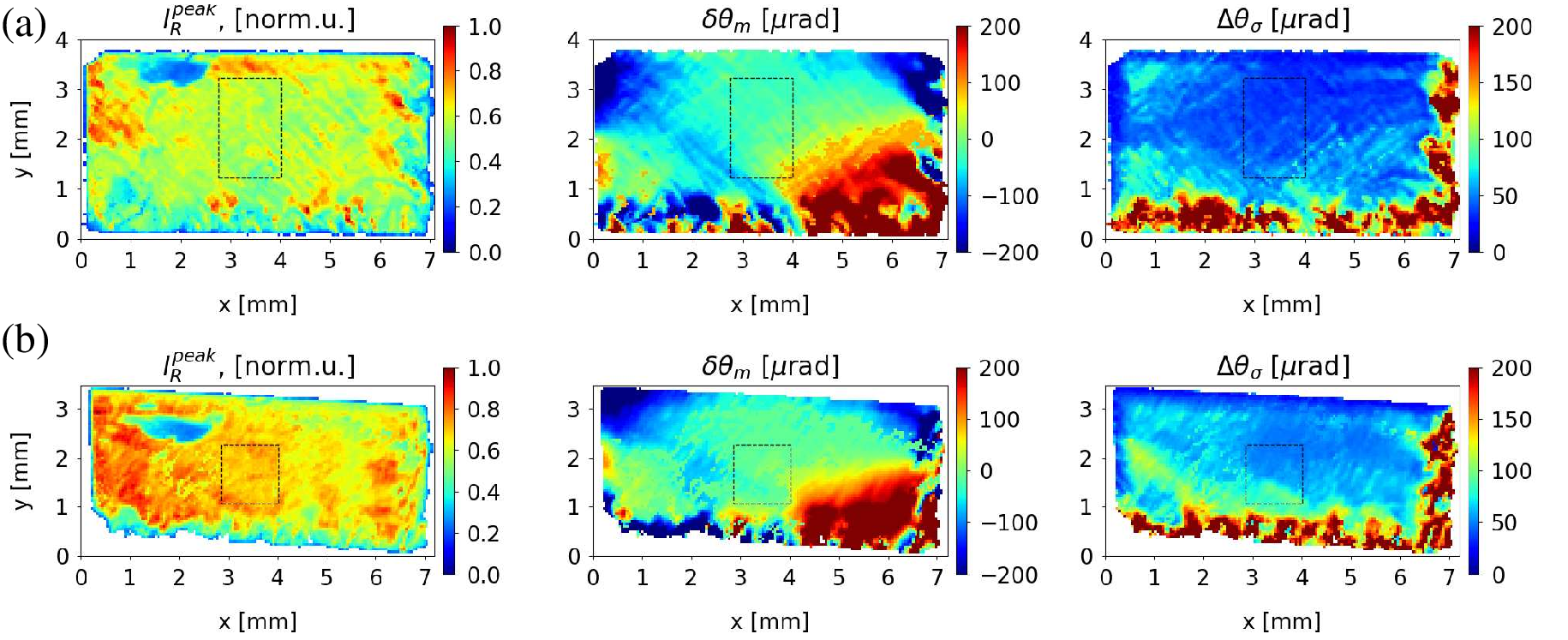} 
\end{figure*}

\begin{figure*}
\caption{Rocking curve topographs of plate CVD-3 in the region probed with the 1$\times$1 mm$^2$ beam 
in the beam expanding geometry (a) and in the beam compressing geometry (b) representing maps of the reflected intensity at the curve's peak value 
normalized by the maximum value observed ($I^{peak}_R$), the curve's peak position ($\delta \theta_m$) and the curve's width as a standard deviation 
of a Gaussian approximation ($\Delta \theta_{\sigma}$). The colorbar on the $\delta \theta_m$ and $\Delta \theta_{\sigma}$ topographs are in units of $\mu$rad.
} 
\includegraphics[scale=0.89]{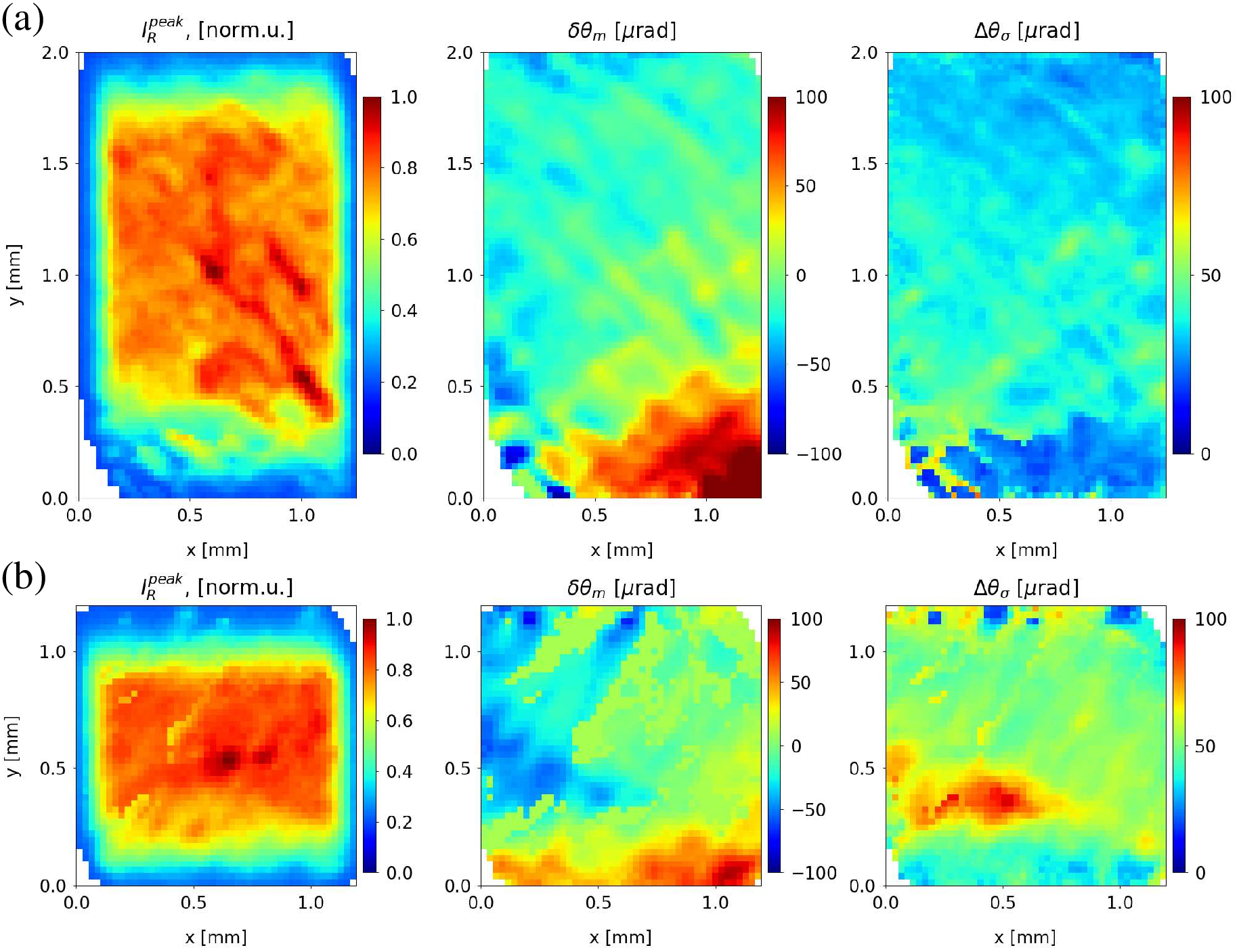} 
\label{fig:rct_wr_cvda}
\end{figure*}

Analysis of rocking curve topographs of plate CVD-3 offers an insight on the most likely scenario.
The topographs of the entire plate CVD-3 in the beam expanding geometry and in the beam compressing geometry
are shown in Fig.~\ref{fig:rct_cvda}(a) and Fig.~\ref{fig:rct_cvda}(b) respectively. The probed region is outlined with a dashed box. 
Rocking curve topographs of the probed region are shown in Fig.~\ref{fig:rct_wr_cvda}(a) and (b), again, in the beam expanding and the beam compressing geometry, respectively. 
The distributions of the reflected intensity ($I^{peak}_R$) is rather homogeneous in the two cases. 
The lower portion of the $\delta \theta_m$ topographs shows similar ($\simeq$~100~$\mu$rad) deviations from the average value represented by zero.
Also, the reflected intensity in this region is reduced below 50\%, which indicates that this distortion, present in both cases, can't cause the discrepancy between 
the two sets of parameters. A region with elevated levels ($\simeq$~100~$\mu$rad) is observed on the $\Delta \theta_{\sigma}$ topograph corresponding to the beam compressing geometry while it is absent in the beam expanding geometry. This is the source of additional broadening of the corresponding rocking curve which was matched with theory using greater values of $\Delta_0$ and $t_0$. Figure~\ref{fig:setup} indicates that the region originates from the lower portion of the crystal, 
which was not illuminated in the beam expanding geometry. Thus, inhomogeneity of the mosaic block structure is a plausible explanation for the observed mismatch. 

All of the studied plates show variations in the model parameters between the two geometries. The analysis of the rocking curve topographs indicates that
inhomogeneities of the crystal in the lateral directions are the likely sources of this variation. This conclusion is further supported by additional measurements
performed using a similar double-crystal setup operating at a photon energy of 8~keV at 1-BM beamline of the Advanced Photon Source. The description and discussion of the results of these measurements for plate CVD-3 are provided in the supporting information. More importantly, the results indicate that for a given crystal volume 
illuminated with X-rays a unique set of model parameters can be used to describe reflectivity at different photon energies. 
Thus, properly chosen mosaic block thickness and the degree of angular misorientation can yield quantitative agreement in absolute X-ray reflectivity of CVD diamond
crystals, which exhibit strong extinction effects for hard X-rays. The mosaic block thickness $t_0$ can be reinterpreted as a characteristic size of a domain over which X-rays are scattered coherently by the crystal lattice \cite{Authier}. This simple picture does not reveal the very details of the crystal microstructure. Nevertheless, it provides a convenient framework for characterization of imperfect crystals as reflectors of X-rays. 

\newpage 
\clearpage
\section{Conclusions}
In summary, absolute X-ray reflectivity of Laue reflections in CVD diamond single crystals was measured in the double-crystal non-dispersive 
setting with an asymmetric beam conditioner crystal. In these measurements rocking curve topography was performed simultaneously on the probed crystal regions. 
Contrary to the case of Bragg geometry Laue geometry yields a manageable X-ray beam footprint on the entrance surface of the crystal for the most efficient 
low-index reflections with shallow Bragg angles. This, along with the superior thermal and radiation hardness properties of 
diamond makes diamond in Laue geometry very attractive for applications in high-heat-load X-ray optics at high photon energies.  
The measured reflectivity curves were described in the framework of Darwin-Hamilton formalism using a set of two independent parameters, a specified thickness of a mosaic 
blocks and their average angular misorientation. In deriving the reflecting power per unit thickness, strong extinction effects in diamond were taken into account  
using numerical integration of the zero-absorption reflectivity solution for a block approximated with a parallel plate of the specified thickness. 
Quantitative agreement between the theory and experiment was obtained by optimal choice of the two parameters. 
The results derived for beam expanding and beam compressing asymmetric geometry for the 111 diamond reflections yielded two different sets of parameters for 
each studied plate. Rocking curve topographs reveal that the source of this discrepancy is due to inhomogeneous microstructure of the diamond plates. Different
crystal volumes were illuminated in the two geometries. 
The developed methodology provides a convenient framework for characterization of imperfect crystals as reflectors of X-rays which 
can be used as guidance in crystal selection procedures to maximize the reflected flux. 
In addition, the results of the study offer an insight on an optimal length scales of the diamond's microstructure, which could offer 
optimal performance in X-ray reflector applications.
 
\acknowledgements

We thank A. Macrander for timely allocation of beamtime for 
experiments conducted at the Advanced Photon Source. K. Lang is acknowledged for technical support. 
Use of the Advanced Photon Source was supported by the U. S. Department of Energy, 
Office of Science, Office of Basic Energy Sciences, under Contract No. DE-AC02-06CH11357. 
This work is based upon research conducted at the Cornell High Energy Synchrotron Source (CHESS) 
which is supported by the National Science Foundation under award DMR-1332208.

\end{document}